\documentclass[twocolumns,reprint,superscriptaddress,aps,pre]{revtex4-1}
\usepackage{amsfonts}
\usepackage{amssymb}
\usepackage{amsmath}
\usepackage{upgreek}
\usepackage{bm}
\usepackage{graphicx}
\usepackage{gensymb,color}

\begin{document}

\title{Modelling Mullins Effect Induced by Chain Delamination and Reattachment}
\author{Daoyuan Qian}
\affiliation{Cavendish Laboratory, University of Cambridge, JJ Thomson Avenue, Cambridge CB3 0HE, U.K.}
\affiliation{CAS Key Laboratory of Theoretical Physics, Institute of Theoretical
Physics,Chinese Academy of Sciences, Beijing 100190, China}

\author{Fanlong Meng}
\email{Corresponding author: fanlong.meng@itp.ac.cn}
\affiliation{CAS Key Laboratory of Theoretical Physics, Institute of Theoretical Physics,Chinese Academy of Sciences, Beijing 100190, China}

\begin{abstract}
We propose a continuum theory to model the Mullins effect, which is ubiquitously observed in polymer composites.
In the theory, the softening of the materials during the stretching process is accounted for by considering the delamination of polymer chains from nano-/micro-sized fillers, and the recovery effect during the de-stretching process is due to the reattachment of the polymer chains to  nano-/micro-sized fillers.
By incorporating the chain entanglements, Log-Normal distribution of the mesh size in the network, \emph{etc.}, we can obtain a good agreement between our numerical calculation results and existing experimental data.
This physical theory can be easily adapted to meet more practical needs and utilised in analysing mechanic properties of polymer composites.
\end{abstract}

\maketitle

\section{Introduction}
Polymer composites ~\cite{general_rev0,general_rev1,general_rev2,general_rev3,general_rev5,general_revapp,general_rev4} are mixtures of polymer chains with reinforcing filler particles (also named as `filled rubber elastomer'), and can be further categorised into micro or nano-composites depending on the size of the fillers.
By adding fillers into the polymer network, one can easily adjust material properties such as elasticity, thermal conductivity ~\cite{general_thermal,general_thermal2} and wet grip ~\cite{general_wet,general_wet2,general_wet3}, and can also make the rubber more durable against material fatigue ~\cite{background_fatigue,general_fatigue2,general_fracture}.
Thus, filled rubbers have been utilised in numerous daily and industrial applications including tires, seals, medical equipments, \emph{etc}~\cite{general_rev0,general_rev4,general_app,general_revapp}.

A pertinent phenomenon of such polymer composites is Mullins effect: when the material is first stretched and then de-stretched, the de-stretch stress-strain curve lies below the stretch curve and there is a residual strain of the material, and it is called as `ideal' Mullins effect if the stress achieved during the re-stretch coincides with the first de-stretch curve.
On the other hand, it is called as `non-ideal' if the re-stretch curve is somewhere in between the first stretch and the first de-stretch curve, \emph{i.e.}, there is partial recovery up to the maximum strain ever attained, and the stress-strain response resembles a virgin material upon further stretching when exceeding the maximum strain ever reached in the deformation history \cite{background_rev,background_mullins}.
Several mechanisms have been proposed for understanding Mullins effect, such as bond rupture \cite{background_bond}, molecules slipping \cite{background_slip}, filler rupture \cite{background_filler} and disentanglement \cite{background_disent}, but a consensus has yet to be reached due to the absence of direct experimental evidence.

\begin{figure}[htb]
\includegraphics[width=3in]{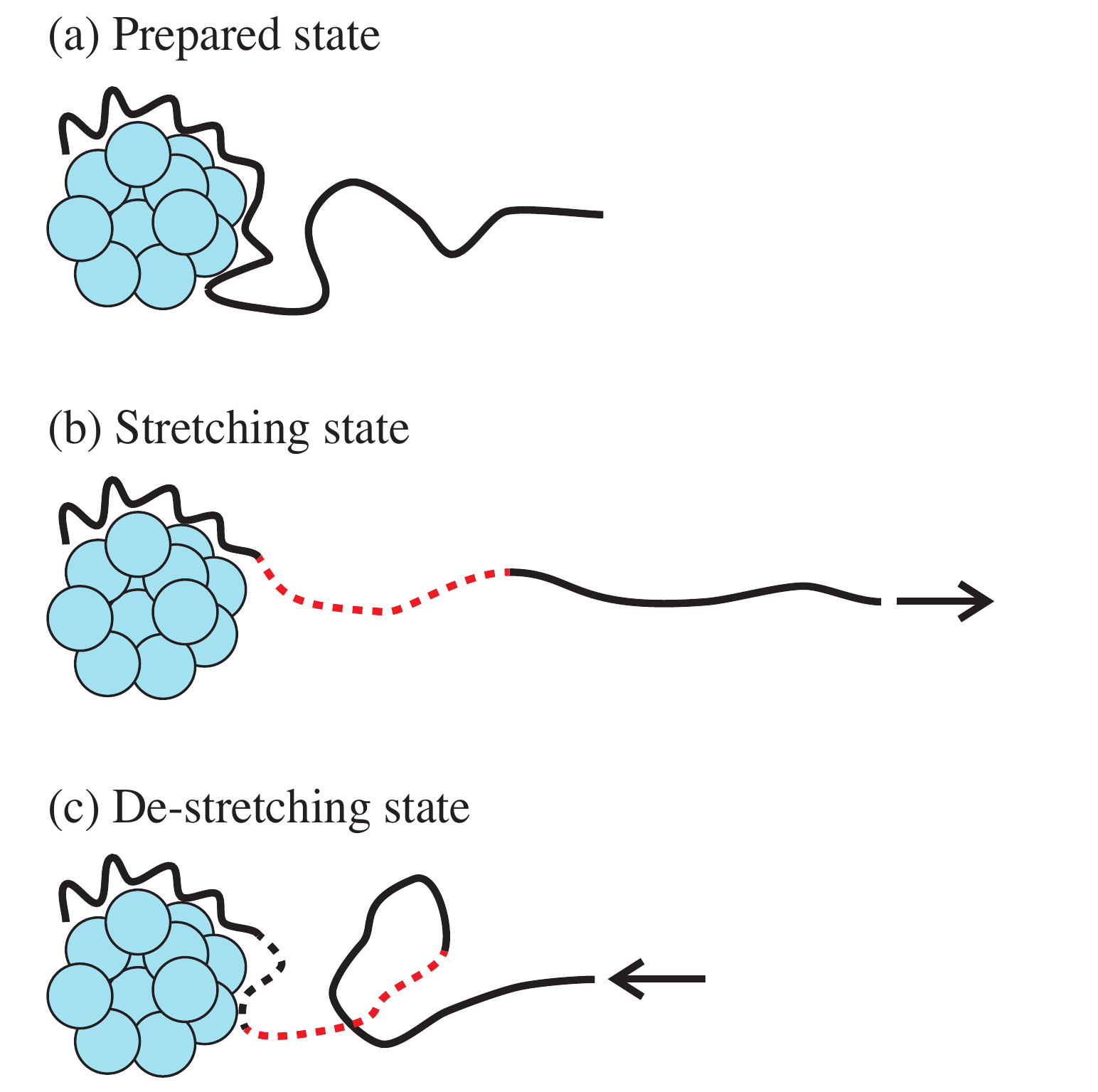}
\caption{Illustration of delamination-induced Mullins effect.
A polymer chain (black) is partly attached to a filler aggregate (blue circles) in the prepared state (a), and partial delamination occurs upon stretching (b), leading to softening.
The delaminated portion is highlighted with red dotted line.
When the material is de-stretched as in (c), part, but not all, of the delaminated chain reattaches back to the aggregate and the softening effect is partially recovered. The portion of reattached chain is indicated with black dotted line.}
\label{illust}
\end{figure}

A recent experiment has provided new insights into Mullins effect \cite{exp_anisotropy}: it was shown that under cyclic loading conditions, the polymer chain anisotropy increased together with Mullins softening, indicating that reversible chain delamination may account for the observed Mullins effect. An illustration of this proposed mechanism is shown in Figure \ref{illust}. Furthermore, no significant change in aggregate size was detected, casting doubt on whether aggregate interactions were indeed important.

In the present work, we propose a continuum model of the energy density function to incorporate the chain delamination and the chain reattachment process, and demonstrate its merit by explaining Mullins effect within our framework. We start with the ideal Mullins effect, showing how delamination alone accounts for softening and residual stretch; then we introduce more practical details including chain reattachment, the distribution of chain length, chain entanglements, \emph{etc.}, from which the non-ideal Mullins effect is recovered, and a comparison of the numerical calculation to experimental data is given.

\section{Ideal Mullins effect}

\subsection{Gaussian chain case}

Suppose that a polymer chain bridging two neighboring filled aggregates consists of $n$ segments, which is not a constant but instead a function of ever experienced deformation,
such that $n\equiv\eta\cdot n_{0}$, with $\eta$
representing the relative elongation of the chain and $n_{0}$ as the initial segment number before any deformation. In the Gaussian limit \cite{general_Boyce},
by denoting $\mathbf{r}$ as the chain end-to-end vector and $l$ as the segment size,
the probability density of finding such a chain with given $\mathbf{r}$ is
\begin{equation}
P(\mathbf{r})=\left(\frac{3}{2\pi \eta n_{0}l^{2}}\right)^{\frac{3}{2}}\exp\left(-\frac{3\mathbf{r}^{2}}{2\eta n_{0} l^{2}}\right),
\end{equation}
and the corresponding stretch ratio of the chain is defined as $\lambda\equiv |\mathbf{r}|/\sqrt{n_{0}}l$.

For a polymer network composed of the above Gaussian chains, which is deformed with the stretching ratios along three orthogonal directions as $\lambda_{1}$, $\lambda_{2}$ and $\lambda_{3}$, respectively,
its energy density function, $F$, can be obtained by using the three-chain model \cite{general_3ch,general_meng,general_Boyce, Meng2016},
\begin{equation}
F(\lambda_{1,2,3},\eta_{1,2,3})=
\frac{1}{2}Nk_{\textmd{B}}T\left(\frac{\lambda_{1}^{2}}{\eta_{1}}+
\frac{\lambda_{2}^{2}}{\eta_{2}}+\frac{\lambda_{3}^{2}}{\eta_{3}}\right)+\textmd{...}
\end{equation}
where $N$ is the number density of chains, $k_{\textmd{B}}$ is the Boltzmann factor and $T$ is temperature.
This energy expression is also known as the neo-Hookean model.
The omitted terms are independent of $\lambda_{1,2,3}$ and drop out during stress calculations.

By uniaxially stretching the material in the $i=3$ direction with the stretch ratio as $\lambda_{3}$, then there is  $\lambda_{1}=\lambda_{2}=1/\sqrt{\lambda_{3}}$ by considering the incompressibility of the material, \emph{i.e.}, $\lambda_{1}\lambda_{2}\lambda_{3}=1$.
For simplicity, we will replace $\lambda_{3}$ with $\lambda$ in the following discussions.
By assuming there is no chain delamination in the $i=1,2$ directions,
\emph{i.e.}, $\eta_{1}=\eta_{2}=1$, the tensile stress can be obtained as, after dropping the subscripts:
\begin{equation}
\sigma(\lambda,\eta)=\frac{\partial F}{\partial \lambda}=
Nk_{\textmd{B}}T\left(\frac{\lambda}{\eta}-\frac{1}{\lambda^{2}}\right).
\end{equation}
The residual strain after the de-stretching can be easily obtained by setting the above tensile stress to zero, giving
\begin{equation}
\varepsilon_{\textmd{residual}}=\lambda_{\textmd{residual}}-1=\sqrt[3]{\eta}-1,
\end{equation}
which is positive for $\eta>1$.

For demonstrative purposes, we here introduce a toy model for $\eta$:
\begin{enumerate}
\item the changing rate of $\eta$ with respect to the stretch: $\mathrm{d}\eta/\mathrm{d} \lambda=k$ for $\lambda\geq\lambda_{\textmd{max}}$,
    where $\lambda_{\textmd{max}}$ is the maximum stretch ever attained in the past deformation history and $k$ denotes the delamination rate;
\item $\mathrm{d}\eta/\mathrm{d} \lambda=0$ for $\lambda<\lambda_{\textmd{max}}$.
\end{enumerate}

\begin{figure}[htb]
\centering
\includegraphics[width=3in]{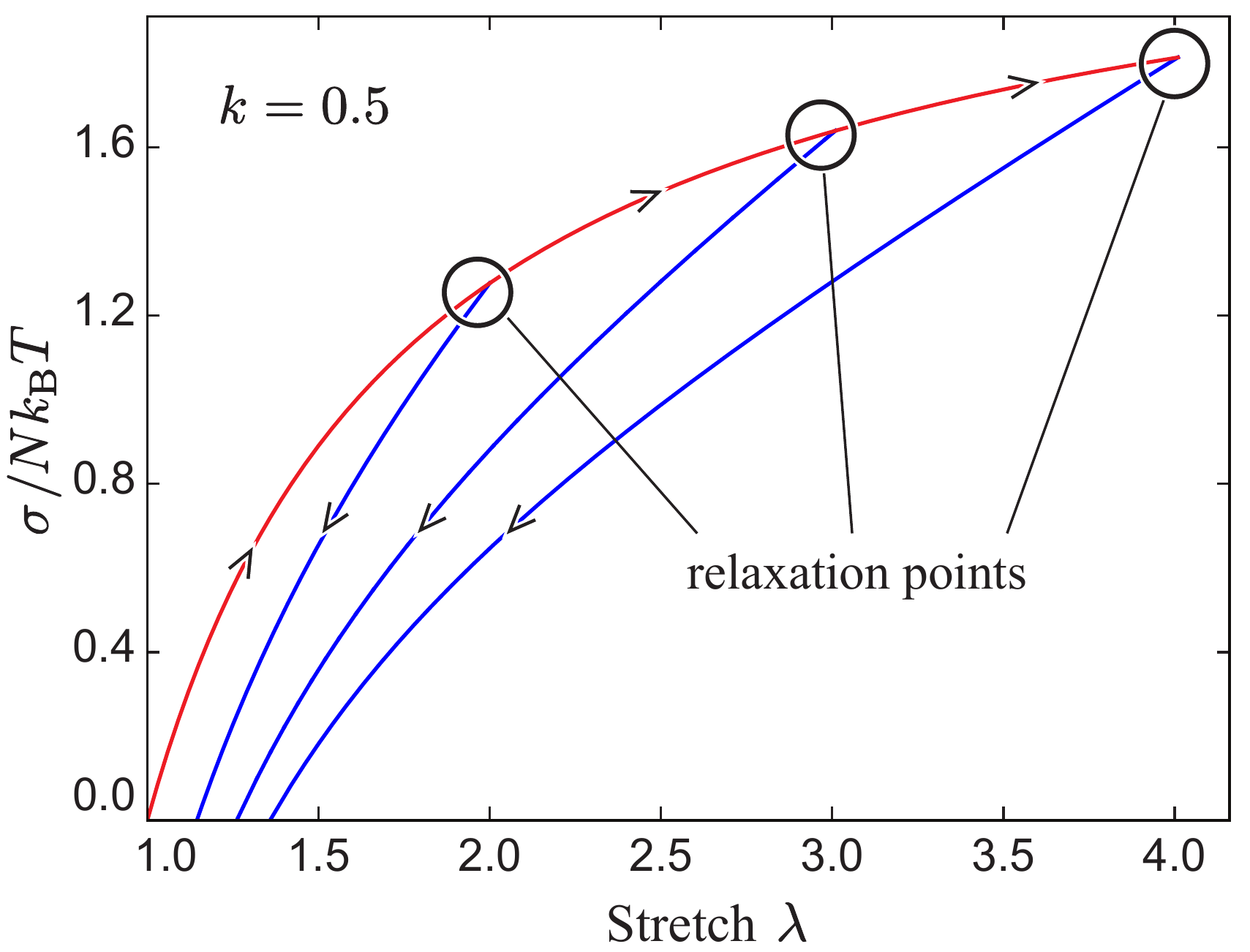}
\caption{Stress-stretch response of the Gaussian chain system. Red and blue lines denote the stress-stretch relations during the stretching process and the de-stretching process, respectively. }
\label{gauss}
\end{figure}

The stress-stretch response is shown in Figure \ref{gauss}. In the Gaussian limit, the stress-stretch response of a chain with constant chain length approaches a linear trend at large stretches and the gradient drops with larger relative elongation, leading to decreasing gradients of the de-stretching curves at relaxation points (circles in Figure \ref{gauss}). This is in stark contrast to the almost-vertical gradients for relaxation stress-stretch curves seen in experiments \cite{background_rev,damage_swell,damage_CDM,network_break} and an important reason is that when a chain is close to the point of delamination, the physical end-to-end distance might be the same order as the contour length of the polymer chain; in this case, Langevin chain statistics \cite{background_kuhn,general_Boyce}, rather than the Gaussian one, is needed.

\subsection{Langevin chain case}

In the non-Gaussian limit, we adopt the Langevin chain statistics treatment \cite{general_Boyce,background_kuhn} under the three-chain framework. The energy of a chain with the end-to-end length $r$ and the segment number $n_{0}$ is
\begin{equation}
W_{\textmd{L}}(r,n_{0})=\sqrt{n_{0}}k_{\textmd{B}}T\left[\frac{r}{\sqrt{n_{0}}l}
\beta+\sqrt{n_{0}}\ln\left(\frac{\beta}{\sinh(\beta)}\right)\right],
\label{energy1}
\end{equation}
with $\beta\equiv\mathcal{L}^{-1}(\frac{r}{n_{0}l})$ as the inverse Langevin function, which in our numerical calculation is approximated by the R. Jedynak form \cite{general_invL}: $\mathcal{L}^{-1}(x)\approx x(3.0-2.6x+0.7x^2)/[(1-x)(1+0.1x)]$
with a maximum fractional error of 1.5\%.
By using the earlier prescription in the Gaussian chain case and replacing $n$ and $r$ with $\eta$, $n_{0}$ and $\lambda$, we arrive at
\begin{equation}
W_{\textmd{L}}(\lambda,\eta,n_{0})=\sqrt{n_{0}}k_{\textmd{B}}T
\left[\lambda\beta+\eta\sqrt{n_{0}}
\ln\left(\frac{\beta}{\sinh(\beta)}\right)\right],
\label{Langevin}
\end{equation}
with $\beta=\mathcal{L}^{-1}\left(\frac{\lambda}{\eta\sqrt{n_{0}}}\right)$. Note that in the small stretch limit, equation \eqref{Langevin} reduces to the Gaussian limit:
\begin{equation}
\lim\limits_{\lambda/\sqrt{n_{0}}\rightarrow0}W_{\textmd{L}}[\lambda,\eta,n_{0}]
=\frac{3}{2}k_{\textmd{B}}T\cdot \frac{\lambda^{2}}{\eta}.
\end{equation}

With the three chain model, we can obtain the energy density function:
\begin{equation}
\begin{split}
F&(\lambda_{1,2,3},\eta_{1,2,3},n_{0})=\\
&\frac{Nk_{\textmd{B}}T}{3}\sqrt{n_{0}}\sum_{i=1}^{3}
\left[\lambda_{i}\beta_{i}+\eta_{i}\sqrt{n_{0}}
\ln\left(\frac{\beta_{i}}{\sinh(\beta_{i})}\right)\right].
\label{3ch}
\end{split}
\end{equation}
Then the tensile stress of a uniaxially stretched material with the stretch ratio $\lambda$, can be obtained as:
\begin{equation}
\begin{split}
\sigma&(\lambda,\eta,n_{0})\equiv\frac{\partial}{\partial \lambda}F(\lambda,\eta,n_{0})\\
&=\frac{Nk_{\textmd{B}}T}{3}\sqrt{n_{0}}
\left[\mathcal{L}^{-1}\left(\frac{\lambda}{\eta\sqrt{n_{0}}}\right)-\frac{1}{\sqrt{\lambda^{3}}}\mathcal{L}^{-1}\left(\frac{1}{\sqrt{\lambda n_{0}}}\right)\right].
\label{3f}
\end{split}
\end{equation}

\begin{figure}[htb]
\centering
\includegraphics[width=3in]{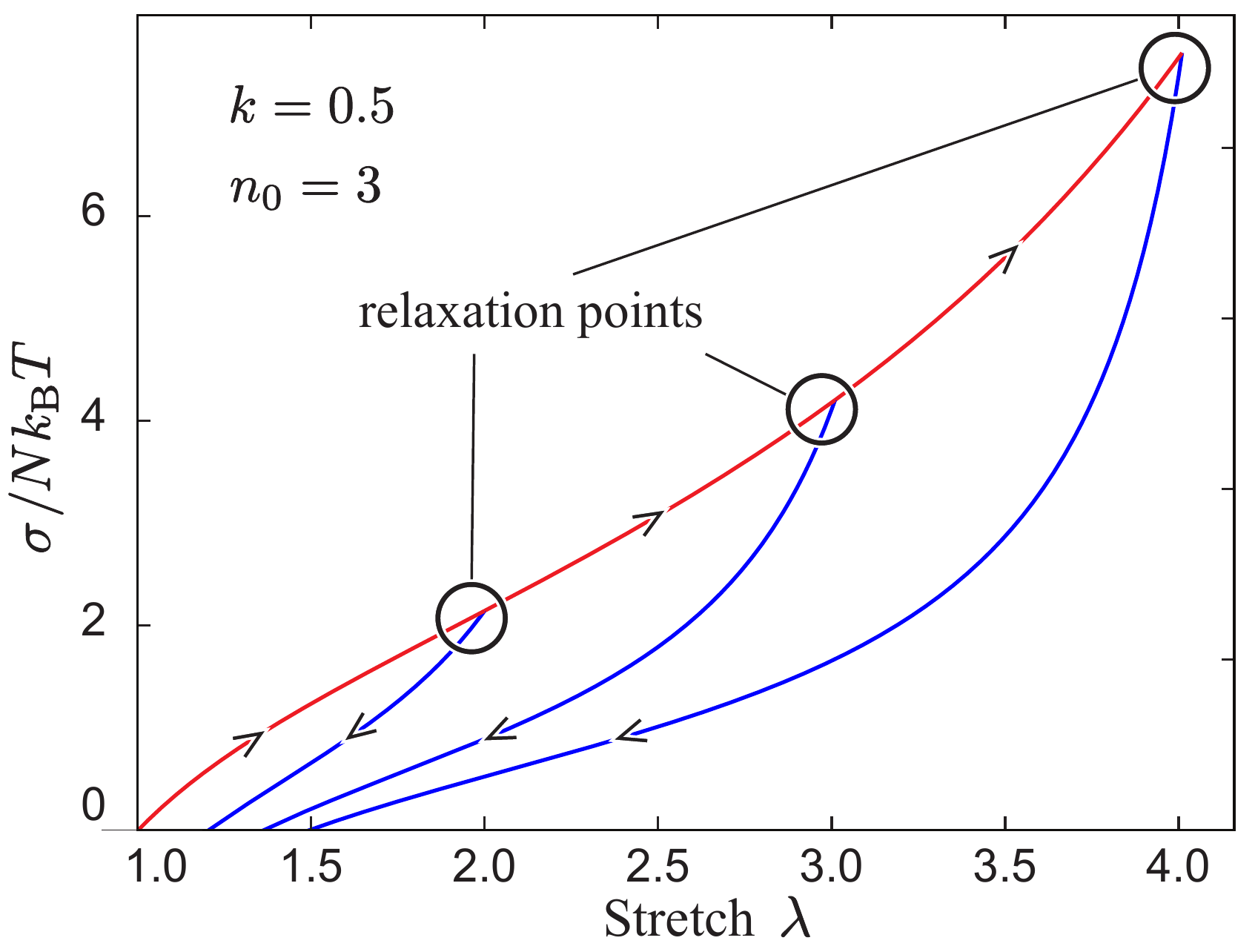}
\caption{Stress-stretch responses of the Langevin chain system. Red and blue lines denote the stress-stretch relations during the stretching process and the de-stretching process, respectively. }
\label{langevin}
\end{figure}

Using the toy model of $\eta$ as introduced in the Gaussian chain case, the ideal Mullins effect with Langevin chain statistics is shown in Figure \ref{langevin}. The Langevin treatment leads to a sharp increase in stress when the chain end-to-end distance approaches the chain contour length and this has indeed reproduced the desired strong non-linearity at relaxation points.

In this section, we have addressed how chain delamination can induce ideal Mullins effect, and the more practical case, non-ideal Mullins effect, will be discussed in details in the next section.

\section{Non-ideal Mullins effect}
In order to give a more realistic description of Mullins effect, in this section we will incorporate chain-length statistics, chain reattachment process, and chain entanglements into the model.

\subsection{Chain-length distribution}

The Langevin force model is highly non-linear in the moderate-to-large stretch range [evident in the $(1-x)$ denominator in $\mathcal{L}^{-1}(x)$] so we cannot simply assume all chains in the rubber have the same length. Here a log-normal distribution of the chain segment number in the prepared state (before any deformation), $n_{0}$, is assumed:
\begin{equation}
P(n_{0})=\frac{Z}{\sigma n_{0}}\exp\left[-\frac{(\ln n_{0}-\mu)^{2}}{2\sigma^{2}}\right]
\end{equation}
with $\mu$ and $\sigma$ as the mean value and the standard variance of $\ln n_{0}$, respectively, and $Z$ as the normalisation factor. In addition,
we denote $n^{\textmd{min}}_{0}$ as the minimal number of chain segments,
the significance of which will be explained in the next sub-section.
Note that the elongation factor $\eta(\lambda)$ will be different for chains with different initial lengths $n_{0}$, and this dependence is denoted by $\eta(\lambda,n_{0})$. Having these in mind we can re-write the energy density function and stress as
\begin{equation}
\begin{split}
F_{\textmd{total}}(\lambda)&=\int_{n^{\textmd{min}}_{0}}^{\infty} P(n_{0})F[\lambda,\eta(\lambda,n_{0}),n_{0}]\mathrm{d}n_{0}\\
\sigma_{\textmd{total}}(\lambda)&=\int_{n^{\textmd{min}}_{0}}^{\infty} P(n_{0})\sigma[\lambda,\eta(\lambda,n_{0}),n_{0}]\mathrm{d}n_{0}.
\label{total}
\end{split}
\end{equation}
Computational details of the integral can be found in Appendix A.
Here we define the elastic modulus, $W_{0}\equiv ZNk_{\textmd{B}}T/3$, as a fitting parameter in the following discussions.

\subsection{Chain delamination during stretching}
Evolution of the chain length distribution has been discussed in network models \cite{network_break, network_stats} and the main idea is that for a single non-Gaussian chain attached to a filler surface, there exists a maximum force $f_{\textmd{max}}$ beyond which the chain will break off from the filler aggregates. Here we argue instead that delamination starts when the force exerted on the chain reaches $f_{\textmd{max}}$, and stops only when the force is below $f_{\textmd{max}}$. The physical significance of $f_{\textmd{max}}$ can be thought of as the bonding force between a polymer segment and the aggregate surface with $f_{\textmd{max}}\cdot l$ as the bonding energy.
The evolution of $\eta$ can be worked out as a function of $\lambda$ in the following manner.

The entropic force exerted by a Langevin chain with the end-to-end length as $r$ and the segment number as $n_{0}$ is \cite{general_Boyce}:
\begin{equation}
f(r,n_{0})=\frac{k_{B}T}{l}\mathcal{L}^{-1}\left(\frac{r}{n_{0}l}\right).
\end{equation}
By incorporating the relative elongation, $\eta(\lambda,n_{0})=n/n_{0}$ to allow chain elongation via delamination, the condition for non-delamination of the chain is given by:
\begin{equation}
\frac{k_{B} T}{l}\mathcal{L}^{-1}\left[\frac{\lambda}{\eta(\lambda,n_{0})\sqrt{n_{0}}}\right]\leq f_{\textmd{max}},
\end{equation}
which can be re-expressed as,
\begin{equation}
\eta(\lambda,n_{0})\geq\frac{\lambda\nu}{\sqrt{n_{0}}},
\label{eta}
\end{equation}
with $1/{\nu}\equiv\mathcal{L}(\frac{f_{\textmd{max}}l}{k_{B}T})$.
In the prepared state without any deformation, both $\lambda$ and $\eta$ are equal to one, and in this case equation \eqref{eta} reduces to $n_{0}\geq\nu^{2}$.
Then one can easily interpret $\nu^{2}$ as the minimum chain length in the material at the prepared state before any deformation,
so we here define $n^{\textmd{min}}_{0}\equiv\nu^{2}$ and re-write equation \eqref{eta} as
\begin{equation}
\eta(\lambda,n_{0})\geq\lambda\sqrt{\frac{n^{\textmd{min}}_{0}}{n_{0}}}\equiv
\eta_{\textmd{d}}(\lambda,n_{0}).
\end{equation}
In other words, when a chain with $n_{0}$ segments is stretched by $\lambda$, its $\eta$ will remain the same as long as $\eta\geq\eta_{\textmd{d}}$, and $\eta$ will become $\eta_{\textmd{d}}$ once the condition $\eta\geq\eta_{\textmd{d}}$ is not satisfied.

\begin{figure}[htb]
\centering
\includegraphics[width=3in]{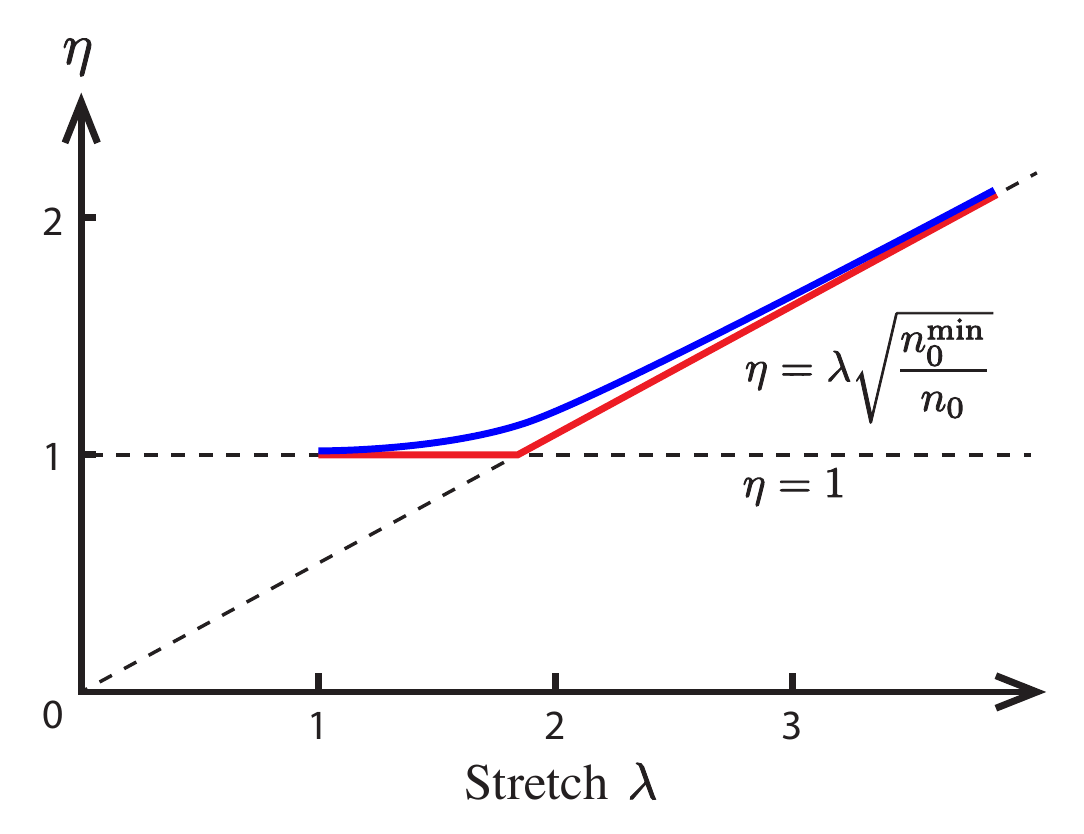}
\caption{Evolution of $\eta$ with $\lambda$ starting from $(\eta=1, \lambda=1)$. Red solid line: evolution with no smoothening; blue solid line: evolution with smoothening.}
\label{delam}
\end{figure}

In practice, the exact force at which a chain delaminates from an aggregate surface may vary if considering the angle between the chain and normal direction of the aggregate surface, and the non-affine nature of the material.
As a result, a smoothening scheme in varying $\eta$ is incorporated, which is shown in Figure \ref{delam}. The degree of smoothening is given by a rate parameter $k_{\textmd{d}}$ and the $\eta$ evolution is governed by:
\begin{equation}
\begin{cases}
\eta_{\textmd{ideal}}=\eta_{d}+ \sqrt{\frac{n^{\textmd{min}}_{0}}{n_{0}}}/k_{\textmd{d}}\\
\mathrm{d}\eta/\mathrm{d} \lambda = 0 \qquad\qquad\qquad~~~~~\ \textrm{if}\ \eta>\eta_{\textmd{ideal}}\\
\mathrm{d}\eta/\mathrm{d} \lambda = k_{\textmd{d}}\cdot (\eta_{\textmd{ideal}}-\eta)\qquad\textrm{if}\ \eta<\eta_{\textmd{ideal}}\\
\end{cases}
\end{equation}
for which more details are shown in Appendix B.

\subsection{Chain reattachment during relaxation}

As the polymer network is relaxed from the stretched state (de-stretching process), the chains start coiling and some polymer segments can come close to the aggregate surface, re-forming polymer-aggregate bonds and resulting in partial recovery of the softening effect.
We propose that the process of chain reattachment can only happen when the current stretch ratio during the de-stretching process is smaller than the residual stretch of the microcell (microcells have different residual stretches, depending on the initial chain length): the three-chain micro-cell has to be compressed, in short. Recalling equation \eqref{3f} and setting the stress to be less than zero, we can obtain
\begin{equation}
\eta(\lambda,n_{0})\leq\frac{\lambda}{\sqrt{n_{0}}\mathcal{L}\left[\frac{1}{\sqrt{\lambda^{3}}}\mathcal{L}^{-1}\left(\frac{1}{\sqrt{\lambda n_{0}}}\right)\right]}\equiv\eta_{\textmd{r}}(\lambda,n_{0}).
\end{equation}

Re-attachment commences when $\eta>\eta_{\textmd{r}}$, and here we assume the rate of change of $\eta$ is proportional to the difference between the current value $\eta$ and the ideal value $\eta_{\textmd{r}}$,
\begin{equation}
\mathrm{d}\eta/\mathrm{d} \lambda=k_{\textmd{r}}\cdot (\eta_{\textmd{r}}-\eta)\label{re} \qquad \textmd{if} \ \eta>\eta_{\textmd{r}}
\end{equation}
with $k_{\textmd{r}}$ the reattachment rate. The ideal Mullins effect has no reattachment and $k_{\textmd{r}}^{(\textmd{ideal})}=0$, while $k_{\textmd{r}}>0$ for non-ideal effects.

\subsection{Entanglement correction}

The final piece added to our model is a correction term arising from entanglement effects to address deviations at small $\lambda$. Here we borrow the entanglement terms from the general constitutive model \cite{Xiang_2018,Xiang_2019,Xiang_2020},  which has energy density and stress:
\begin{equation}
\begin{split}
W_{\textmd{e}}&=G_{\textmd{e}}\left(\frac{1}{\lambda}+2\sqrt{\lambda}\right)\\
\sigma_{\textmd{e}}&=G_{\textmd{e}}\left(-\frac{1}{\lambda^{2}}+\frac{1}{\sqrt{\lambda}}\right)
\end{split}
\end{equation}

The modulus $G_{\textmd{e}}$ is subject to damage of the form \cite{Xiang_2020}:
\begin{equation}
G_{\textmd{e}}=G_{\textmd{e0}}\exp\left[-\frac{k_{\textmd{e}}}{2}\left(\sqrt{I_{1}^{\textmd{max}}/3}-1\right)\right]
\end{equation}
with $G_{\textmd{e0}}$ the initial entanglement modulus, $k_{\textmd{e}}$ the damage rate, and $I_{1}^{\textmd{max}}$ the maximum value of the first invariant of the Cauchy-green tensor ever attained in the material's history. An illustration of how the entanglement correction contributes to the total stress can be found in Appendix C.

\subsection{Numerical results}
\begin{figure}[htb]
\centering
\includegraphics[width=3in]{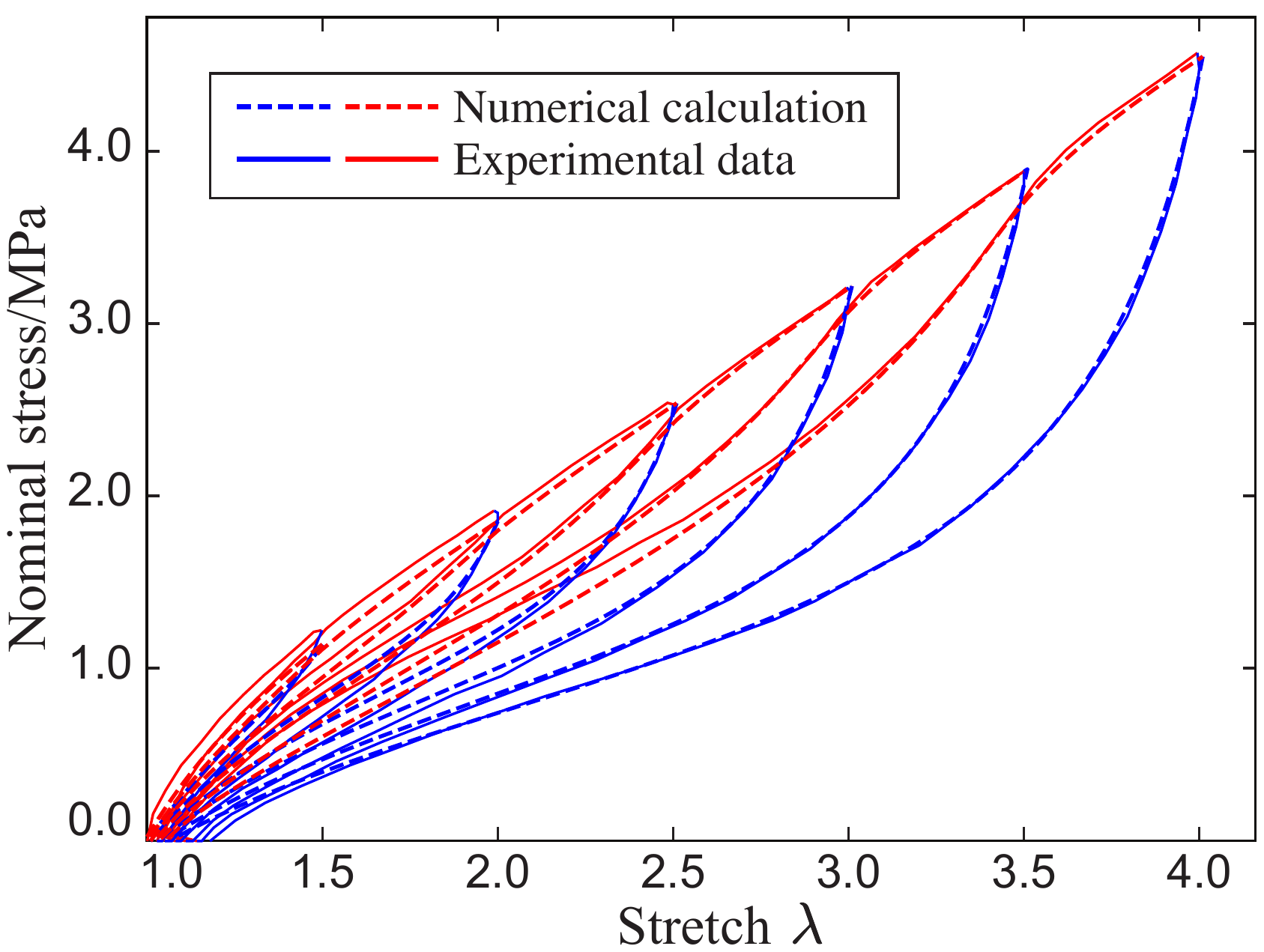}
\caption{Comparison of our model (dashed lines) to experimental data (solid lines) \cite{damage_swell}. Red and blue lines denote the stress-stretch relations during the stretching process and the de-stretching process, respectively. }
\label{Malaysia_full}
\end{figure}

\begin{figure}[htb]
\centering
\includegraphics[width=3in]{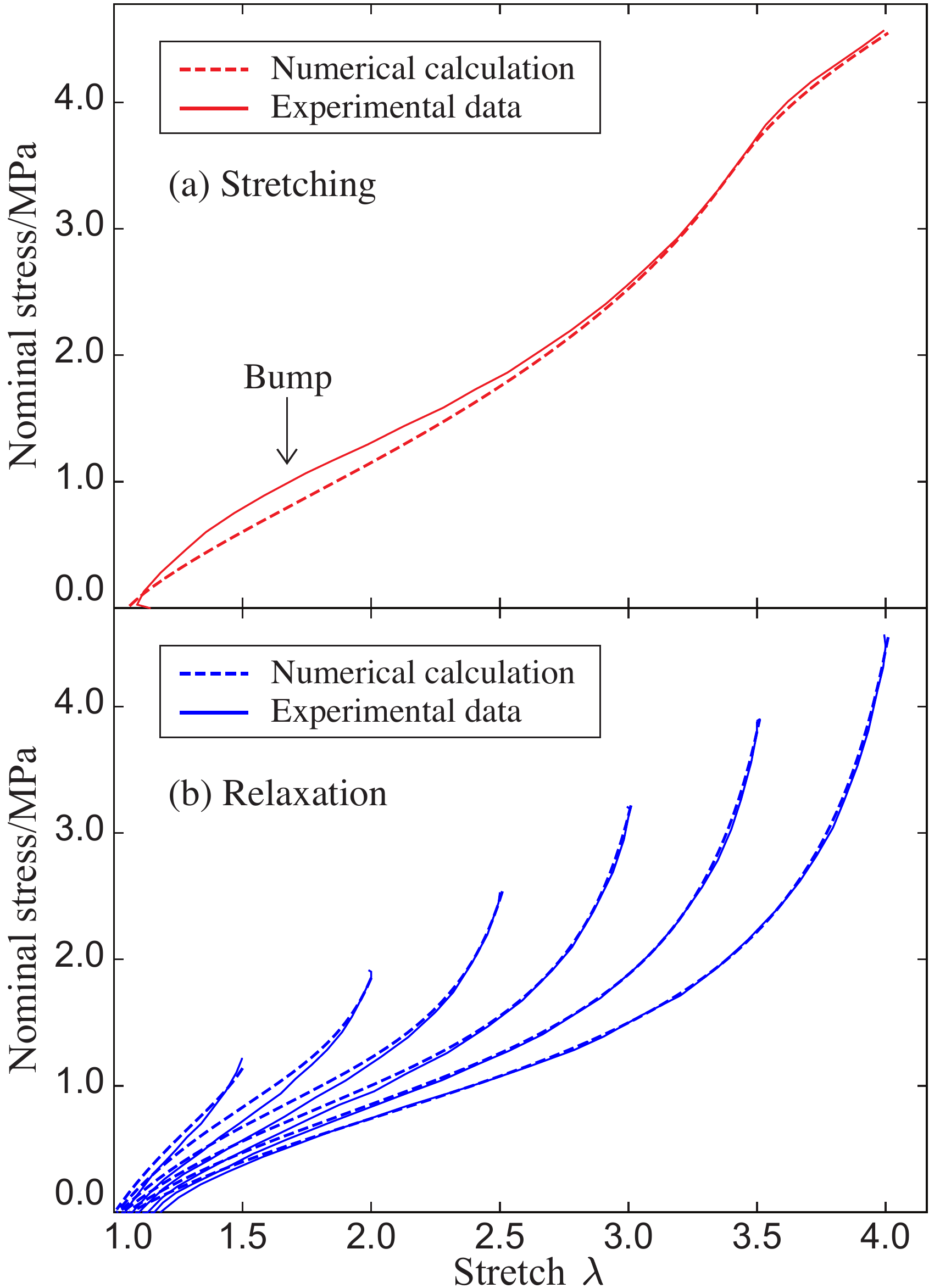}
\caption{Stress-stretch relations for (a) stretching process and (b) de-stretching process. }
\label{devi}
\end{figure}

We fit our model to experimental data extracted from previous works \cite{damage_swell} as an example (table \ref{parameters} shows parameters used) and Figure \ref{Malaysia_full} shows how our model compares to experiment. A good agreement is achieved, especially for large stretch ratios, say $\lambda > 2.5$, with strong initial non-linear responses for relaxation across all stretches as desired.
Apart from the satisfactory fitting between our theory and the experiments, there are, however, also observable deviations at small stretches.
For example, in Figure \ref{devi}(a), the experimental data shows a `bump' compared to our theoretical result during the stretching process.
One way to incorporate such `bumpy' behaviours is by the addition of aggregate elasticity \cite{network_MIT}, which is not considered in the current work to keep the relative transparency of the theory.

\begin{table}[htb]
\centering
\caption{Fitting parameters}
\begin{tabular}[t]{llc}
\hline
Parameter&Description&Value\\
\hline
\hline
$W_{\textmd{0}}$&Elastic modulus&0.072 (MPa)\\
$n^{\textmd{min}}_{0}$&Minimum chain length&1.17\\
$k_{\textmd{d}}$&Delamination rate&6\\
$k_{\textmd{r}}$&Reattachment rate&8\\
$\mu$&Log-Normal parameter 1&2.8\\
$\sigma$&Log-Normal parameter 2&1.6\\
$G_{\textmd{e0}}$&Entanglement modulus&1.1(MPa)\\
$k_{\textmd{e}}$&Entanglement damage&2.5\\
\hline
\end{tabular}
\label{parameters}
\end{table}

\section{Conclusion}
To conclude, we have developed a physical model to understand the non-ideal Mullins effect ubiquitously observed in filled rubber elastomers, by incorporating both chain delamination from and chain reattachment to the filled aggregates. Good agreements can be obtained between the model calculation and experimental data. We believe this portable theory can be easily adapted to meet more practical needs and utilised in analysing mechanic properties of polymer nano-/micro-composites.

\begin{appendix}

\section{Integration method}

The integration in equation \eqref{total} extends to infinity and traditionally a large-n cutoff will be used in the integration so that the integration range covers 95\% of the chains. In the actual computation, however, we observe that the chains assume the Gaussian behaviour at large $n$ and delamination only occurs for $n_{0}$ up to $n_{0}\sim\lambda_{\textmd{max}}^{2}\sim16$, so the stress contribution becomes independent of $n_{0}$ and can be taken out of the integral.

Expanding $\mathcal{L}^{-1}(x)$ to third order we have $\mathcal{L}^{-1}(x)\approx3x+0.1x^{2}+1.09x^{3}...$and since $x\sim1/\sqrt{n}$, with a cut-off $n_{\mathrm{max}}=50$ and truncating the series to first order, the fractional error in $\mathcal{L}^{-1}(x)$ is $\sim(0.1x+x^{2})/3\sim1.2\%$, similar to the accuracy achieved by the R. Jedynak approximation. The nominal stress calculation can then be written as
\begin{equation}
\begin{split}
\sigma_{\textmd{total}}(\lambda)\approx&\int_{n_0^{\mathrm{min}}}^{50} P(n_{0})\sigma[\lambda,\eta(\lambda,n_{0}),n_{0}]\mathrm{d}n_{0}\\
+&Nk_{\textmd{B}}T\left(\lambda-\frac{1}{\lambda^{2}}\right)\cdot \int_{50}^{\infty} P(n_{0})\mathrm{d}n_{0},
\end{split}
\end{equation}
and the integration from $50$ to $\infty$ can be computed with Mathematica. The first integration has to be done numerically and 10,000 points are used in our simulation.

\section{Smoothening of $\eta$ evolution}

The smoothening procedure for $\eta$ during delamination is similar to equation \eqref{re}. We start by solving the following simplified model for $y(x)$ and $y_{\textmd{ideal}}(x)$

\begin{equation}
\begin{cases}
y_{\textmd{ideal}}(x)=\alpha x \\
\frac{\mathrm{d}y(x)}{\mathrm{d}x} = \beta[y_{\textmd{ideal}}(x)-y(x)] \\
y(0)=0
\end{cases}
\end{equation}
which has solution $y(x)=\alpha x-\frac{\alpha}{\beta}(1-e^{-\beta x})$. Note that the solution asymptotically approaches $\alpha x - \frac{\alpha}{\beta}$ instead of $y_{\textmd{ideal}}(x)$, so $y_{\textmd{ideal}}(x)$ has to be shifted upwards by $\frac{\alpha}{\beta}$ for $y(x)$ to approach $\alpha x$. Going back to our $\eta-\lambda$ system we can make the identification $(x,y)=(0,0)\rightarrow(\lambda,\eta)=\left(\sqrt{\frac{n_{0}}{n^{\textmd{min}}_{0}}},1\right)$, $\alpha\rightarrow\sqrt{\frac{n^{\textmd{min}}_{0}}{n_{0}}}$ and $\beta\rightarrow k_{d}$, so the evolution law has the following form:
\begin{equation}
\begin{cases}
\eta_{\textmd{ideal}}=\eta_{r}+ \sqrt{\frac{n^{\textmd{min}}_{0}}{n_{0}}}/k_{\textmd{d}}\\
\mathrm{d}\eta/\mathrm{d} \lambda = 0 \qquad\qquad\qquad\qquad\qquad\ &\textmd{if}\ \eta>\eta_{\textmd{ideal}}\\
\mathrm{d}\eta/\mathrm{d} \lambda = k_{\textmd{d}}\cdot (\eta_{\textmd{ideal}}-\eta) &\textmd{if}\ \eta<\eta_{\textmd{ideal}}\\
\end{cases}
\end{equation}

\section{Entanglement contribution}
\begin{figure*}[htb]
\centering
\includegraphics[width=7in]{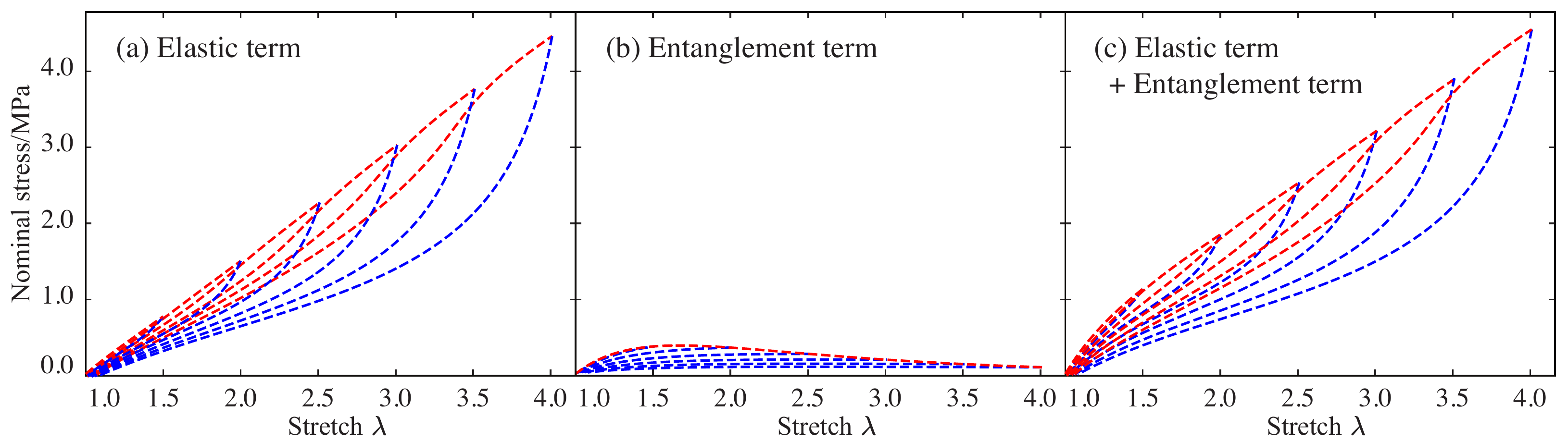}
\caption{Stress-stretch relations for considering only (a) the elastic term and (b) the entanglement term. (c) Stress-stretch relations for adding the contributions of both the elastic term and the entanglement term. Red and blue lines denote the stress-stretch relations during the stretching process and the de-stretching process, respectively.}
\label{compare}
\end{figure*}
Figure \ref{compare} shows separately how the elastic term and entanglement term each contribute to the total stress. Although the fitting parameter for the elastic term is much smaller than the entanglement term, the rapidly-increasing inverse Langevin function meant that the elastic contribution still dominates. With only elastic term [Figure \ref{compare}(a)] the initial stress-stretch response appears linear due to the averaging effect of integrating over all chain lengths, and the entanglement term [Figure \ref{compare}(b)] adds a bump at the start to match the experimental observation.

\end{appendix}

\end{document}